\documentclass[aps,twocolumn,a4paper,showkeys,showpacs]{revtex4-1}
\usepackage{amsmath,amssymb,bm,xcolor,mathrsfs,graphicx,subfigure,texdraw}
\usepackage{mathtools}

\usepackage{color}

\mathtoolsset{showonlyrefs}

\definecolor{ColorMG}	{rgb}{0.000000,0.500000,0.000000}

\begin{document}

\title{Strain intermittency in shape-memory alloys}

\author{Xavier Balandraud$^{1}$, Noemi Barrera$^{2}$, Paolo Biscari$^{2}$,  Michel Gr\'ediac$^{1}$, and Giovanni Zanzotto$^{3}$}

\affiliation{$^1$Clermont Universit\'{e}, Universit\'{e} Blaise Pascal, IFMA, Institut Pascal, UMR CNRS 6602, BP~10448, 63000~Clermont-Ferrand, France\\
$^2$Department of Physics, Politecnico di Milano, Piazza Leonardo da Vinci 32, 20133 Milano, Italy\\
$^3$DPG, Universit\`a di Padova, Via Venezia 8, 35131 Padova, Italy}



\begin{abstract}

We study experimentally the intermittent progress of the mechanically-induced martensitic transformation in a Cu-Al-Be single crystal through a full-field measurement technique: the grid method. We utilize an in-house especially designed gravity-based device, wherein a system controlled by water pumps applies a perfectly monotonic uniaxial load through very small force increments. The sample exhibits hysteretic superelastic behavior during the forward and reverse cubic-monoclinic transformation, produced by the evolution of the strain field of the phase microstructures. The in-plane linear strain components are measured on the sample surface during the loading cycle, and we characterize for the first time the strain intermittency in a number of ways, showing the emergence of power-law behavior for the strain avalanching over almost six decades of magnitude. We also describe the non-stationarity and the asymmetry observed in the forward vs.~the reverse transformation. The present experimental approach, which allows for the monitoring of the reversible martensitic transformation both locally and globally in the crystal, proves useful and enhances our capabilities in the analysis and possible control of transition-related phenomena in shape-memory alloys.

\end{abstract}

\maketitle

\section{Introduction}

Shape-memory alloys (SMAs) are active crystalline materials with desirable mechanical properties, exploited in a number of applications from engineering to medicine, as actuators, sensors and dampeners \citep{olson, otsuka, lexcellent}. In these substances, effects such as super- (or pseudo-)elasticity and shape memory originate from a reversible martensitic transformation triggered by either stress or temperature. These phenomena are strongly affected by the formation of austenite/martensite  microstructures, typically including twins and habit planes in various arrangements which may act as proper avenues for the reversible phase change~\citep{wedge_91, ball_97, sedlak_05, cui_06, BZ_07, stupkiewicz_07, zhang_09, delville_10, Balandraud_Acta_Mat_10, Harrison_10, nature_14}. A number of theoretical approaches have been developed for the analysis of such morphologies, going back to the classical `crystallographic theory of martensite' \citep{Wechsler, Bowles, Wayman, Nishiyama}, based on considerations of kinematic compatibility between phases and variants. This viewpoint has been much developed in recent decades~\citep{James_Acta_Mat_00, book_PZ, Bhattacharya_03, nature_04, science_05, nature_14}.

In parallel, the experimental investigation of SMA microstructure has adopted a variety of means, primarily optical and electron microscopy, originating an extensive literature, see for instance the references in \citep{olson, otsuka, lexcellent}. Developments in the techniques of full-field measurements have recently also enabled the experimental analysis of the microstructural strain field over extended regions of a sample's surface. Moir\'e interferometry and digital image correlation
are typical examples of such techniques, which collect information on the response of the tested SMA samples by distinguishing the different phases through their strain amplitude~\citep{Zhang_J_Phys_97, Efstathiou_Scripta_Mat_08, Merzouki_Mech_Mat_10, Daly_Exp_Mech_09, Sanchez-Arevalo_Mater_Characterization_09}.
Infrared thermography has also been employed to measure the inhomogeneous temperature distributions associated with the phase change in SMAs, giving relevant information on the features of the martensitic transformation~\citep{X1, X2, X3, X4, Delpueyo_MSEA_11, Delpueyo_Mech_Mater_12, Pieczyska_13}.

The spatial distribution of the phases and variants over a sample can also be deduced through the grid method \cite{Badulescu_Meas_Sci_Technol_09}, wherein this information is derived by analyzing the deformation of a periodic grid deposited on the sample, obtaining the strain levels which are generally different among the phases and their variants. A main advantage of the grid method is that it offers a good compromise between strain resolution and spatial resolution \cite{Badulescu_Meas_Sci_Technol_09}. However, this full-field measurement technique have not yet been employed to investigate phase-change intermittency in memory materials. This inhomogeneity is a fundamental feature of the martensitic transformation in SMAs, wherein under a smooth thermal or mechanical driving, the strain progresses through a set of discrete avalanche-like events, corresponding to transitions between neighboring metastable states. Such effects have been previously studied through the measurement of calorimetric and/or acoustic signals~\citep{Vives_94, vives95, Carrillo_Phys_Rev_B_97, Carrillo_Phys_Rev_Lett_98,  perez_04a, Vives_Phys_Rev_B_09, niemann14, toth_14}. These techniques have revealed many aspects of the bursty character of the martensitic transformation, with events exhibiting, in many cases, power-law size distributions, possibly after training~\citep{PTZ_07}. This links reversible SMA martensites to other material systems exhibiting intermittent dynamics and scale-free behavior, in magnetism, superconductivity, brittle fracture, crystal plasticity \citep{kardar, fisher, alava, zaiser, sornette}, in turn framing them within the background of a wide variety of complex systems in nature exhibiting avalanche-mediated behavior, as in turbulence, earthquakes, computer or social networks, and financial markets \citep{sornette, bak, newman}.

While recent investigations have examined quantitatively some spatial aspects in the evolution of martensitic transition phenomena \citep{Harrison_10, Vives_Phys_Rev_B_11, niemann14}, there is a lack of systematic quantitative information on the strain events derived from the analysis of evolving strain maps during the phase change in SMAs. In the present study we used the grid method to perform the first full-field investigation of the intermittent progress of the microstructural strain field in a Cu-Al-Be single crystal across the mechanically-induced martensitic transformation. This was done through uniaxial force-controlled loading tests, during which we have recorded and analyzed the super-elastic and hysteretic stress-strain behavior of the sample, as well as surface strain data in the forward and reverse transition.

Inspired by earlier work on the stress-controlled loading of SMAs in~\citep{Carrillo_Phys_Rev_B_97} (see also \cite{Bonnot07, Vives_Phys_Rev_B_09}), an experimental apparatus based on gravity was used, which applied to the sample a low-rate perfectly monotonic force, controlled by very slow flow water pumps, at almost isothermal conditions. The loading device was designed to be capable of very small force increments, enabling us to highlight the intermittency in the alloy, through the quantitative analysis of the strain field features measured over the sample. In contrast to conventional testing machines, as used for instance in~\cite{Delpueyo_MSEA_11, Delpueyo_Mech_Mater_12}, in the present apparatus the crystal may freely adjust its orientation in relation to the vertical loading direction. A main reason for this was obtaining relatively less complex microstructures, developed in the presence of minimal extraneous effects such as external friction, lattice plastification, or thermal inhomogeneity, to enable us to investigate the strain transformation intermittency occurring in the crystal in its most elementary and basic form.

This paper is organized as follows. Sect.~\ref{section_2} presents the sample under study, the experimental setup, the loading conditions annd the procedure employed for image processing to retrieve strain maps. Sect.~\ref{section_3} presents an analysis of the hysteresis and microstructures deduced from the strain evolution recorded during a full loading test. Sect.~\ref{section_4} discusses the observed intermittent character of the phase transformation. Sect.~\ref{section_5} gives some conclusive remarks. Further details on the experimental apparatus and on data acquisition and treatment are given in the Appendix. The Supplemental Material \cite{online_video} comprises four video files showing different aspects of the experimental results.

\section{Experimental methods}\label{section_2}

\begin{figure}
\begin{center}
\includegraphics[width=\linewidth]{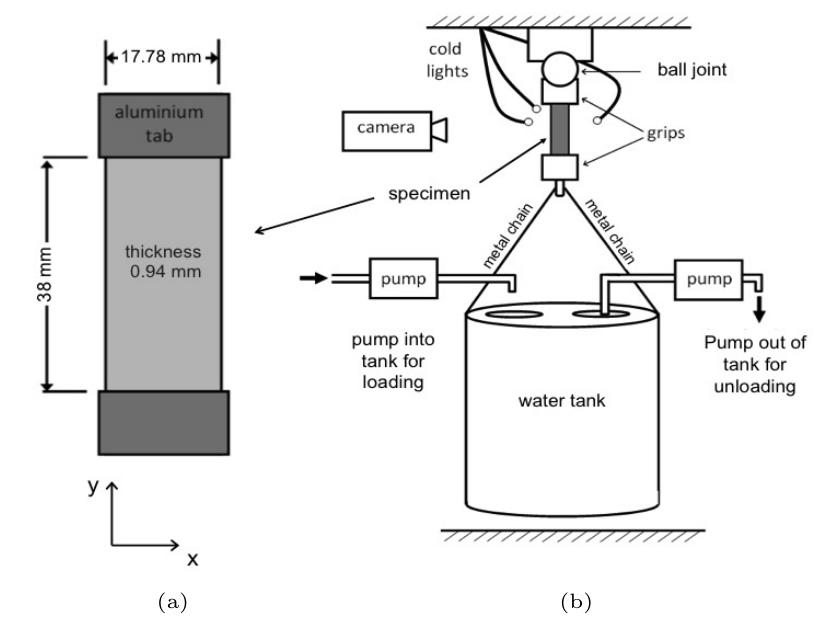}
\end{center}
\caption{Schematic representation of (a) the SMA sample with the gauge region in gray, and (b) the testing device. See the Appendix for details.}
\label{figure_1}
\end{figure}

We tested a Cu\,Al$_{11.4}$\,Be$_{0.5}$ (wt.~\%) single crystal sample, the same used in the loading experiments described in~\citep{Delpueyo_MSEA_11, Delpueyo_Mech_Mater_12}, to which we refer for additional information on the crystallographic properties and orientation of the lattice in the sample (see the Appendix). This SMA, under uniaxial tensile loading, undergoes a reversible martensitic transformation,
from a cubic austenite to a long-period stacking monoclinic martensite with twelve variants, which are all compatible with the austenitic phase \cite{James_Acta_Mat_00, BZ_07}. The two-fold axes in the monoclinic variants originate from the cubic axes in the austenite \cite{book_PZ}.

Fig.~\ref{figure_1}(a) shows the sample geometry. Aluminum tabs were clamped in two pairs of jaws by means of screws onto the ends of the sample, to avoid damage or sliding once positioned in the testing machine. A bidirectional grid (pitch $p= 0.2\,$mm) was transferred using the technique described in~\cite{Piro_Exp_Tech_04} on the monitored gauge region (dimensions $17.78$ mm $\times$ 38 mm respectively along the $x$- and $y$-directions) covering nearly all the exposed surface of the SMA sample, with thickness $0.94$ mm. See the Appendix for more details.

\begin{figure}
\begin{center}
\includegraphics[width=\linewidth]{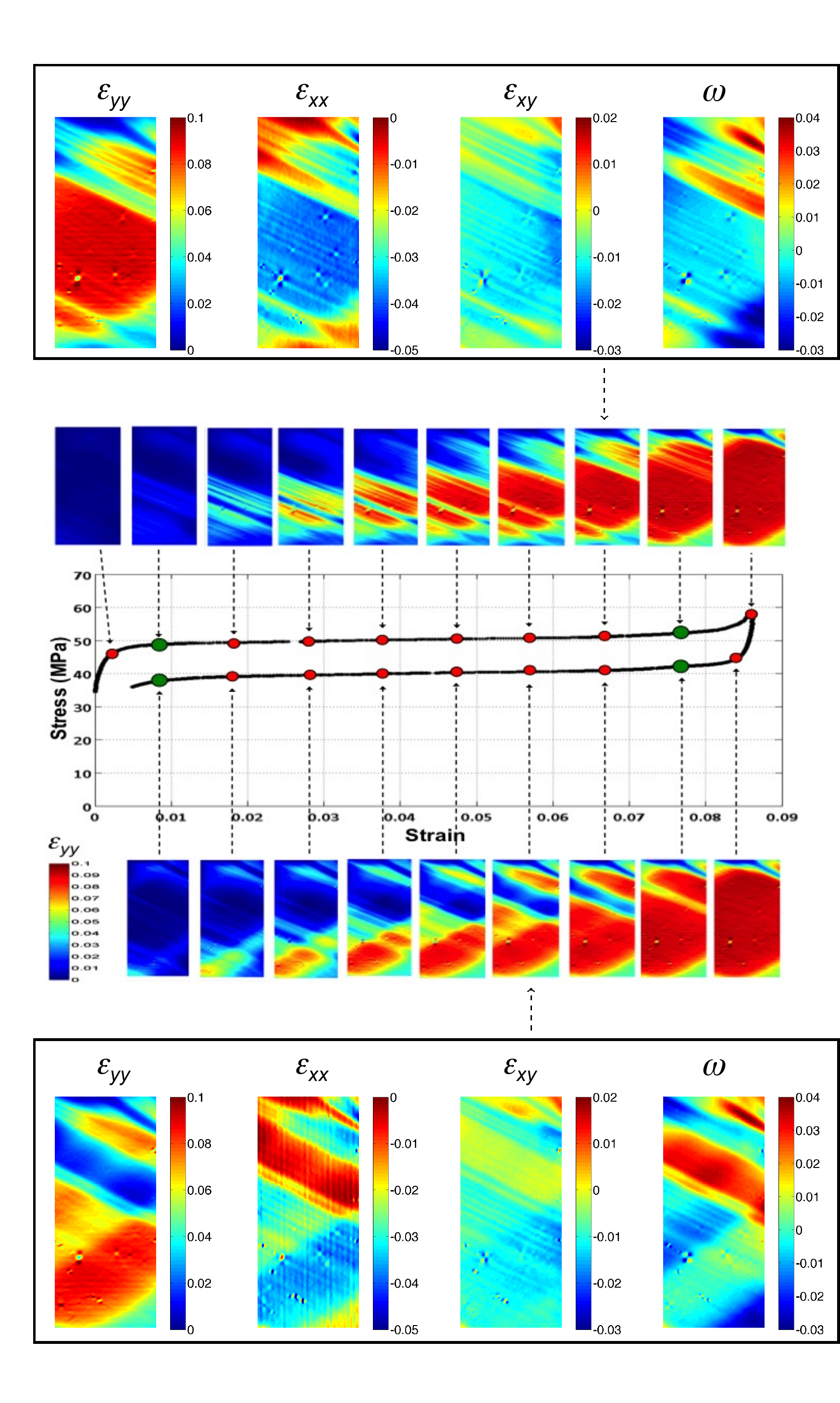}
\end{center}
\caption{(Color online) Stress-strain curve and snapshots of the strain field $\varepsilon_{yy}$ at various stages of the loading-unloading cycle (austenite is in blue, martensite in red). The stress-strain hysteresis regards the average strain $\overline{\varepsilon}_{yy}$ vs.~the average stress $\overline{\sigma}_{yy}$. The green dots mark points on the transformation plateaus for reference in the figures below. Also shown are maps of all the in-plane strain components and of the angle $\omega$, corresponding to the average strain value $\overline{\varepsilon}_{yy}=0.067$ on loading (top), and to $\overline{\varepsilon}_{yy}= 0.057$ on unloading (bottom). Notice that in the $\varepsilon_{xx}$ map the austenite is in red and the martensite in blue. See the supplemental video~\#1~\citep{online_video}.}
\label{hysteresis}
\end{figure}

The loading apparatus is schematically shown in Fig.~\ref{figure_1}(b).~The sample, which before testing was austenitic with some possible residual martensite, was suspended at its top end through a ball joint which allows for any rotations. A tank was hung through a cable to the jaws at the bottom end of the sample. The tank was filled with water during the experiments, enabling us to apply a gravity-load in the vertical direction. Loading and unloading of the tank were controlled by electronic pumps producing a constant and slow flux of water. This allowed for a careful stress control of both the forward and reverse austenite-to-martensite phase change upon loading and unloading, respectively. As transformation stresses depend on the temperature, we ensured the stability of the ambient temperature $T_\text{amb} = 26.8\pm 0.5\,^\circ$C during the entire duration of the test (more than 45\,h). Due to the very slow driving, the testing conditions were almost isothermal. Before using the pumps, a pre-load was applied to the sample to reduce the test duration shortening the first part of the elastic climb where martensitic transformation events are negligible. This gave the reference configuration for the ensuing strain maps. The forward and reverse transformation plateaus were spanned through a slowly-varying, monotonic loading controlled by adjusting the water flow through the pumps.

In what follows we specifically report on a test in which we have observed the whole forward and reverse transformations under these conditions: (a) pre-load of 34.37\,MPa; (b) constant loading rate 1.055\,MPa/h up to 57.29\,MPa (duration: about 22\,h); (c) constant unloading rate $-0.915$\,MPa/h down to 35.95\,MPa (duration: about 23\,h 15\,min). Due to the inherent feedback loop in the control, very slow monotonic rates such as the ones above, obtained with the present device, are not possible with conventional testing machines. As a comparison, in earlier stress-controlled tests, such as in~\citep{Carrillo_Phys_Rev_B_97}, the rate was about 37\,MPa/h in the gauge region of the sample.

We have gathered about 20,000 images of the sample over the 45-hour duration of the test, from which we have determined the spatial distribution of the strain levels on the sample surface by applying suitable grid-image processing. This enabled us to obtain four maps from each image during the loading-unloading cycle, giving the in-plane linear strain fields $\varepsilon_{xx}$, $\varepsilon_{yy}$, $\varepsilon_{xy}$, and the rotation-angle field $\omega$ about the $z$-direction (see Fig.~\ref{figure_1} for the axes orientation). More details on the data acquisition and treatment are given in the Appendix.

\section{Hysteresis and microstructure}
\label{section_3}

\begin{figure}
\begin{center}
\includegraphics[width=\linewidth]{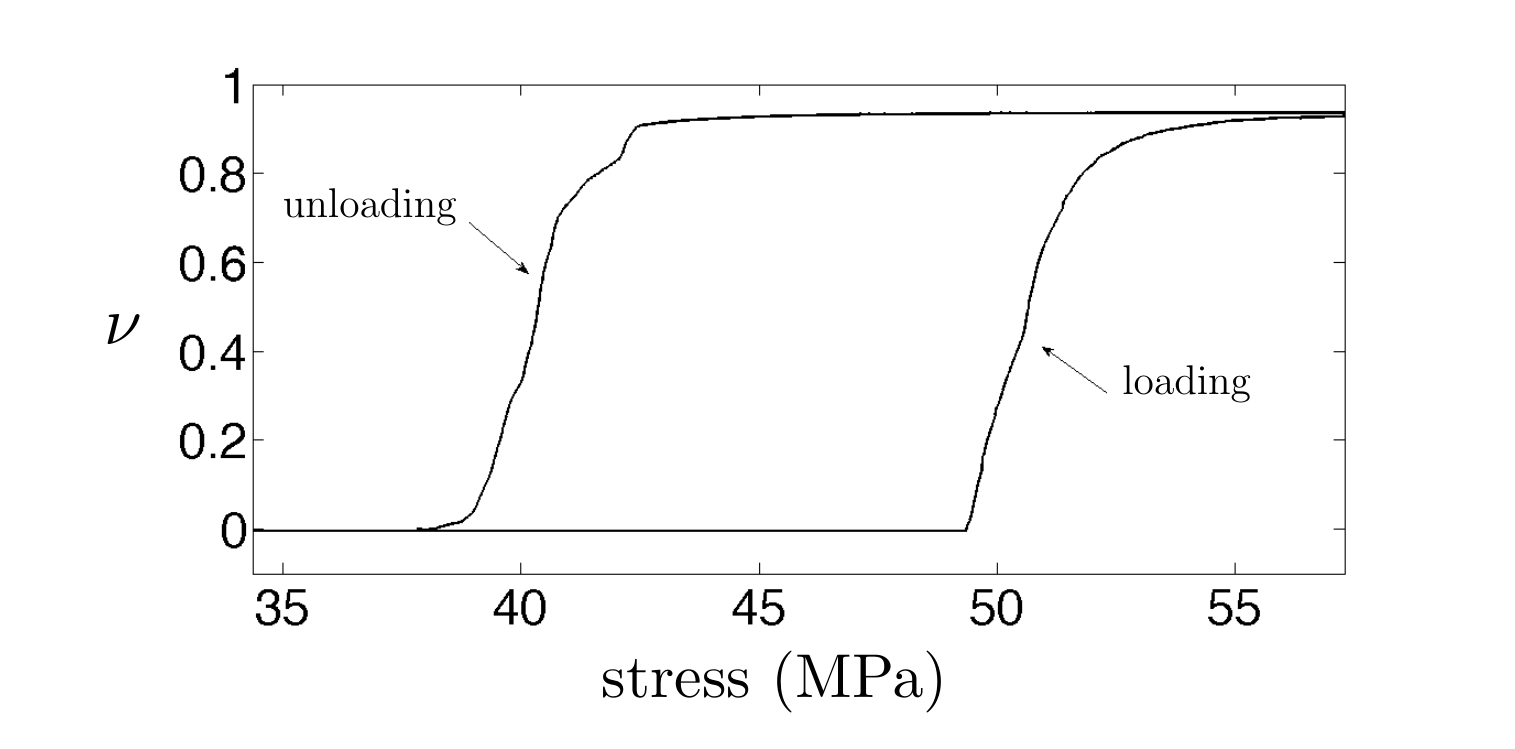}
\end{center}
\caption{Hysteresis in the evolution of the martensitic volume fraction $\nu$ vs.~the average stress $\overline{\sigma}_{yy}$ during the stress-driven forward and reverse transformation in Fig.~\ref{hysteresis}. The fact that $\nu$ does not reach 1 originates from the possible presence of both residual martensite at the lowest loading as well as some residual austenite at the highest loading.}
\label{evolution_nu}
\end{figure}

Fig.~\ref{hysteresis} presents the hysteretic stress-strain relation recorded during the loading-unloading cycle. Further information is given in the supplemental video~\#1~\citep{online_video}. Fig.~\ref{hysteresis} also shows the evolving strain maps through the forward and reverse transformations, where we see that, expectedly, $\varepsilon_{xx} < 0$ and $\varepsilon_{yy} > 0$. Fig.~\ref{evolution_nu} shows the associated hysteresis in the evolution of the martensitic volume fraction $\nu$, defined as the percentage of the sample surface where~$\varepsilon_{yy}$ exceeds the threshold~$\tilde \varepsilon_{yy}=0.05$, which is about half of the maximum local strain value recorded during the test.

\begin{figure}
\begin{center}
\includegraphics[width=7.5cm, height=7cm]{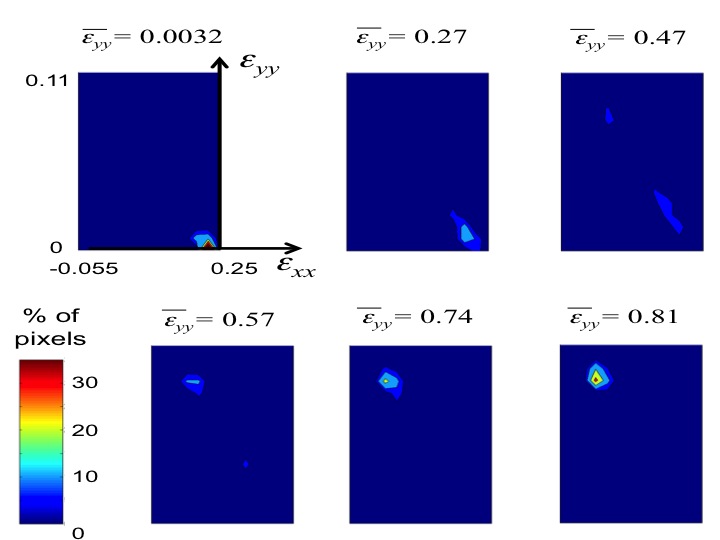}
\end{center}
\caption{(Color online) Strain clustering in the sample during the forward transformation from austenite to martensite. Snapshots for selected growing values of the average strain $\overline{\varepsilon}_{yy}$ are shown. The color bar indicates the fraction of pixels with the given strain values, with grid pitch on the $\varepsilon_{xx}$ and $\varepsilon_{yy}$ axes equal respectively to 0.005 and 0.007. See also the supplemental video~\#2~\cite{online_video}.}
\label{strain_clouds}
\end{figure}

We observe in Fig.~\ref{hysteresis} the development of the strain microstructure as the phase transformation progresses from austenite (blue region in the $\varepsilon_{yy}$ map) towards a unique variant of martensite (red region in the $\varepsilon_{yy}$ map), which is selected among the twelve possible variants due to the suitable alignment of its lattice with the imposed load. The presence of a single variant in this test is indicated by the very homogeneous values for the field $\varepsilon_{yy}$ in the red martensitic domain of the sample (with average approaching $\overline{\varepsilon}_{yy} = 0.09$ near the end of the transformation), and by the fact that the shear component $\varepsilon_{xy}$ has constant sign. The occurrence of a single variant is also suggested by Fig.~\ref{strain_clouds}, showing how the material evolves between strain values mostly clustering at two levels, corresponding to (strained) austenite and to a single (strained) variant of martensite. At the intermediate loading configurations in Fig.~\ref{strain_clouds} a trail is noticed between the pure-phases' strain values. This is due to the elastic deformation of the phases' lattices under while the load is present, as well as to the fact that any small scale phase domains in the sample are below the spatial resolution of the apparatus (see the Appendix), and thus give rise to averaged strain values in the corresponding pixels. The evolution of the material from the austenitic energy well to the martensitic one along the upper transformation plateau can also be viewed in the supplemental video~\#2~\citep{online_video}.

The $\varepsilon_{yy}$ strain maps in Fig.~\ref{hysteresis} show that the forward transformation develops through needle-like bands which progressively enlarge their width and whose length eventually crosses the entire sample. Their inclination agrees with the theoretical estimates \cite{PhDBarrera} based on the compatibility equations giving the habit-planes for this transformation \cite{book_PZ, Bhattacharya_03}. The snapshots in Fig.~\ref{hysteresis} also confirm the existence of relative lattice rotations between regions occupied by different phases, with jumps of the angle $\omega$ from +0.03 to $-$0.03 rad across the austenite-martensite interfaces, again in good agreement with the theoretically computed values \cite{PhDBarrera}.  The $\varepsilon_{yy}$-field in Fig.~\ref{hysteresis} appears more complex along the reverse transformation, possibly involving also a second habit plane \cite{PhDBarrera} (see more details below, and the supplemental videos~\citep{online_video}). The evolution on a loading cycle is thus hysteretic not only in terms of the stress-strain relation, but also in the distinct microstructural morphologies developed in loading vs.~unloading. This might be a factor contributing to a slight mechanical irreversibility in the phase change.

\begin{figure}[t!]
\begin{center}
\includegraphics[width=\linewidth,height=9cm]{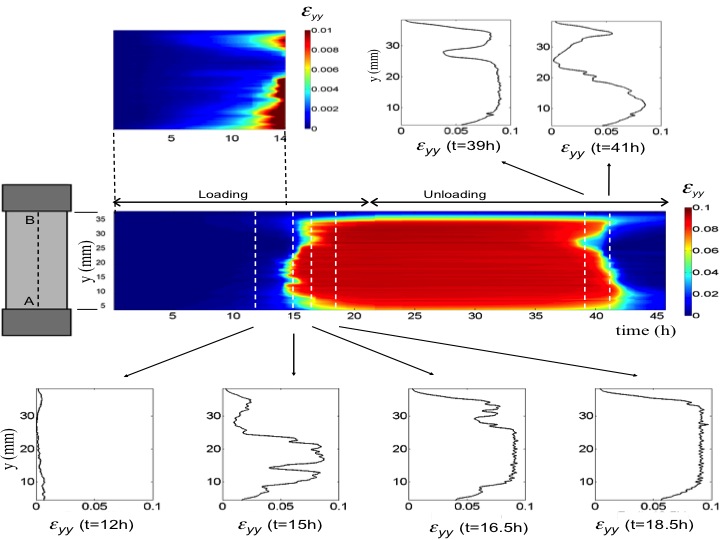}
\end{center}
\caption{(Color online) Time evolution of the strain component $\varepsilon_{yy}$ along the mid-section~AB of the sample. The detail in the upper left corner of the figure shows the evolution of $\varepsilon_{yy}$ in the interval [0\,h, 14\,h] at a magnified strain scale.}
\label{time_evolution_E}
\end{figure}

Fig.~\ref{time_evolution_E} shows, as an example, the time evolution of the strain component~$\varepsilon_{yy}$ along the central section AB of the gauge region. Most of the phase transformation occurs within the time window [14\,h, 19\,h] on loading (the upper plateau in the stress-strain curve of Fig.~\ref{hysteresis}), and [38\,h, 43\,h] on unloading, with a total stress increase/decrease of about 5\,MPa on each plateau. This gives the overall slant of the hysteresis loop in the stress-strain plane in the present test. The upper left corner of Fig.~\ref{time_evolution_E}  shows the evolution of~$\varepsilon_{yy}$ in the interval [0 h, 14 h] at a magnified strain scale. The strain profiles in Fig.~\ref{time_evolution_E} show the evolution of the phase microstructures in Fig.~\ref{hysteresis}, where we also notice that the end pixels A and B and their adjacent regions do not undergo significant strain.

\begin{figure}
\begin{center}
\includegraphics[width=\linewidth, height=8.3cm]{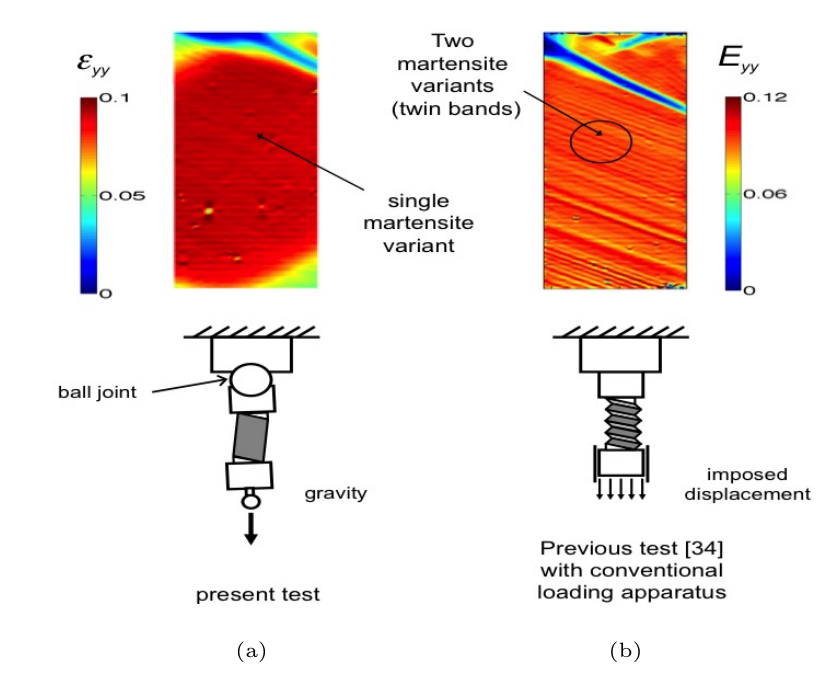}
\end{center}
\caption{(Color online) Comparison of the strain fields obtained in the present loading test (a), and in a test performed on the same sample with a conventional uniaxial loading device in~\citep{Delpueyo_Mech_Mater_12} (b). In both cases the austenite-to-martensite phase transformation is nearly complete: blue and red shades indicate the austenitic and martensitic regions in the sample. $E_{yy}$ and $\varepsilon_{yy}$ denote respectively the Hencky and linear strain components. The faint horizontal lines in the left snapshot are artifacts due to slight pitch variations in the grid, not completely eliminated by the data treatment. The small dots originate from local grid defects.
}
\label{comparison}
\end{figure}

It is interesting to contrast these results with those recorded in earlier tests performed on the same Cu-Al-Be sample by means of a conventional uniaxial loading machine, which also exploited the grid method \citep{Delpueyo_Mech_Mater_12}. Fig.~\ref{comparison} reports the $yy$-strain maps in the two cases, near the end of the transformation when comparable small fractions of austenite remain in the sample (blue regions). The martensitic (red) regions show the different microstructures developed in the two tests. As noted above, the present loading device produced a homogeneous $\varepsilon_{yy}$-field  (Fig.~\ref{comparison}, left), while the conventional machine produced a martensitic zone with two alternating strain levels (Fig.~\ref{comparison}, right), which we interpret as martensite twinning. The development of twin bands is indeed expected in the conventional machine, as the latter does not allow for rotations of the sample, forcing the material to activate more than one martensite variant to accommodate the imposed load. Conversely, a single variant can accommodate the loads in the present testing device, as the latter allows for an overall rotation to accompany the transformation strain in the sample. This helped in producing less complex microstructures, for this first full-field study of strain intermittency in SMAs.

\section{Intermittency}
\label{section_4}

\begin{figure}[h]
\begin{center}
\includegraphics[width=0.47\textwidth]{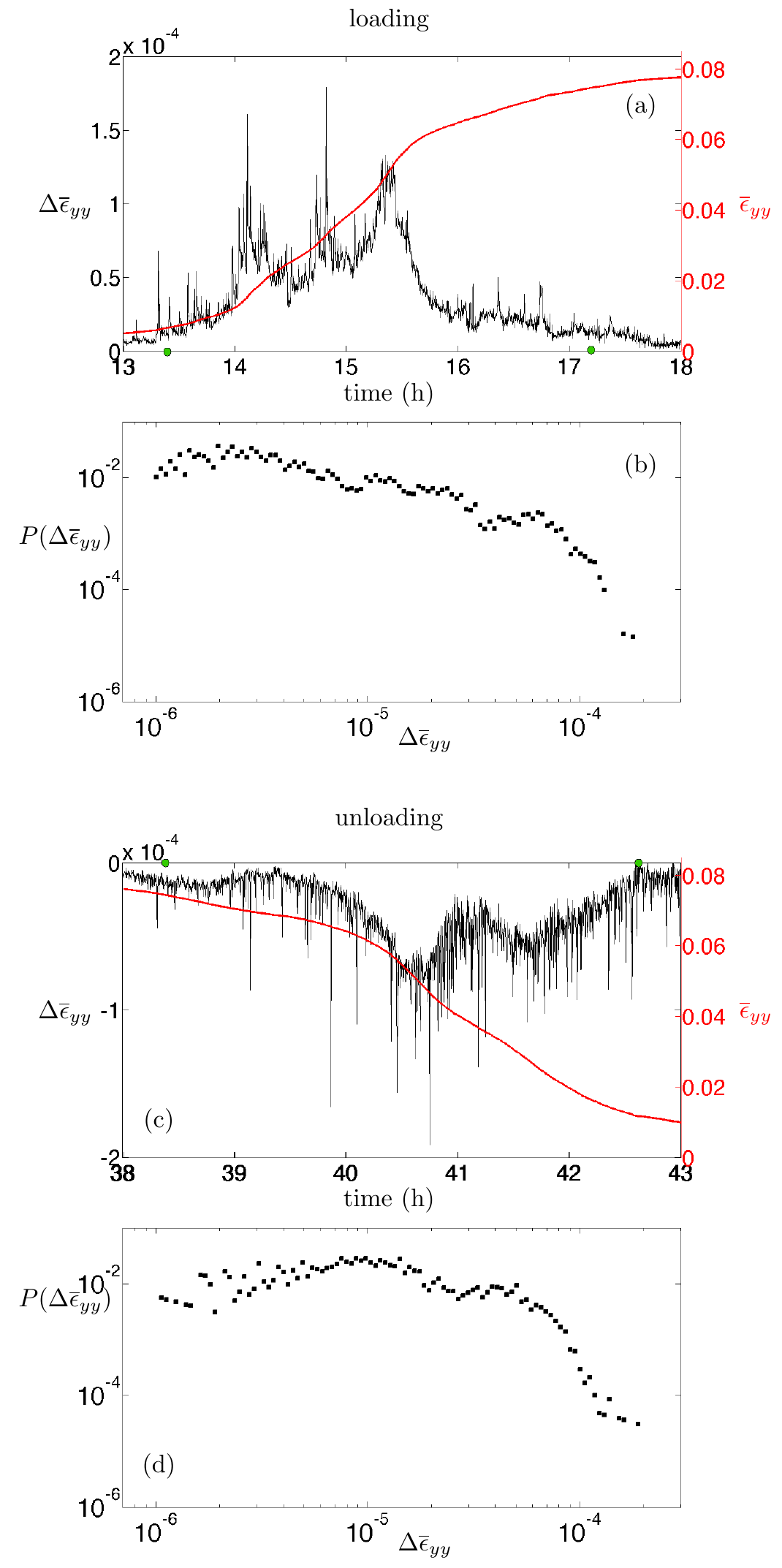}
\end{center}
\caption{
Time evolution of the jumps $\Delta\overline{\varepsilon}_{yy}$ recorded at each consecutive image for the average strain $\overline{\varepsilon}_{yy}$ during loading (a), and unloading. The reference green dots mark the transformation plateaus. Reported in red is the corresponding time evolution of $\overline{\varepsilon}_{yy}$, see also Fig.~\ref{hysteresis}.  The corresponding heavy-tailed probability distributions for $P(\Delta\overline{\varepsilon}_{yy})$ are also shown (logarithmic binning).}
\label{spikes_delta_epsilon_mean}
\end{figure}

We study here various aspects of the non-continuous progress of the observed forward and reverse martensitic transformation. We see in Fig.~\ref{hysteresis} that the global variable $\overline{\varepsilon}_{yy}$ displays a rather continuous evolution at the time scale of the entire test (the same holds for $\nu$ in Fig.~\ref{evolution_nu}). However, Fig.~\ref{spikes_delta_epsilon_mean} shows that the evolution in time of the increments $\Delta\overline{\varepsilon}_{yy}$ obtained by comparing each subsequent image taken during the test, produces a very spiky signal, evidencing a pronounced strain intermittency in the phase-transforming sample under the slowly-varying monotonic load. Fig.~\ref{spikes_delta_epsilon_mean} shows the corresponding heavy-tailed probability distributions $P(\Delta\overline{\varepsilon}_{yy})$ for the $\Delta\overline{\varepsilon}_{yy}$ amplitudes over almost two decades (we threshold the values of $\overline{\varepsilon}_{yy}$ and $\Delta\overline{\varepsilon}_{yy}$ at $10^{-6}$, see the Appendix). Yet another form of hysteresis in the forward vs.~the reverse transformation is highlighted in this way, as we see from Fig.~\ref{spikes_delta_epsilon_mean} that most of the jumps $\Delta\overline{\varepsilon}_{yy}$ are systematically smaller on unloading than on loading. However, the largest events are more numerous in the reverse than in the forward transformation. Analogous behavior is observed for the spiky signal given by the phase-fraction jumps $\Delta\nu$ (not shown). Fig.~\ref{spikes_delta_epsilon_mean} also evidences that the recorded strain intermittency is non-stationary along the transformation cycle, whereby there is more activity in the evolution of $\Delta\overline{\varepsilon}_{yy}$ near the center of both the plateaus of the hysteresis as opposed to their ends (more details on this below).

\begin{figure}
\begin{center}
\includegraphics[width=\linewidth]{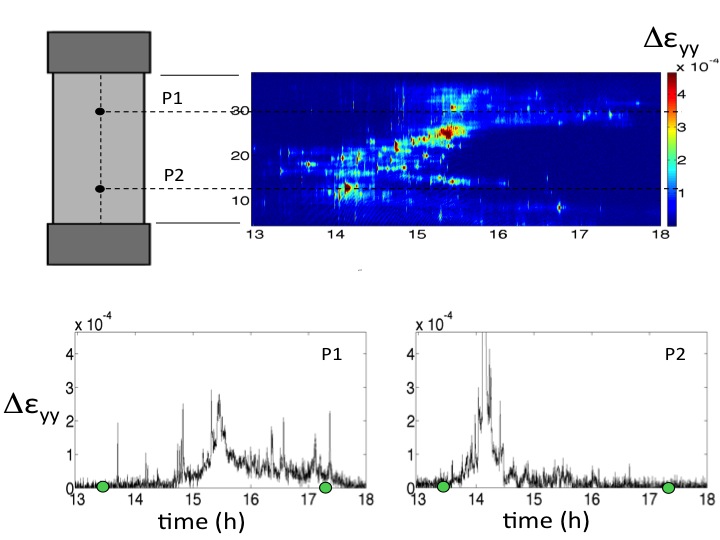}
\end{center}
\caption{(Color online) (a) Evolution in time of the jumps $\Delta\varepsilon_{yy}$, evaluated at each consecutive image, along the mid-section of the sample. See also Fig.~\ref{time_evolution_E}. (b) Time evolution of $\Delta\varepsilon_{yy}$ at two pixels P1 and P2. The time interval is [13\,h; 18\,h], where most of the phase transformation occurred on loading. The reference green dots delimit the transformation strain plateaus.}
\label{local_bursts}
\end{figure}

The intermittent evolution of the global variables such as ($\nu$ and) $\overline{\varepsilon}_{yy}$ originates from the intermittency in the underlying strain field in the sample. The latter is highlighted in Fig.~\ref{local_bursts}, which reports the bursty behavior of the strain increments $\Delta\varepsilon_{yy}$ at some typical locations on the surface (the increments again refer to each consecutive image). The main panel shows the space-time evolution of $\Delta\varepsilon_{yy}$ along the sample mid-section on loading. The lower panels plot the temporal evolution of  $\Delta\varepsilon_{yy}$ at two pixels P1 and P2 of the same section. Strain events occur throughout the forward and reverse phase change in the crystal, with the larger bursts proportionally likely to include phase transformation in the lattice as opposed to purely elastic local deformation. We notice high-intensity (yellow-to-red) localized strain bursts occur throughout the sample, indicating the local evolution of the phase microstructures in the crystal. The largest events in Fig.~\ref{local_bursts} are of the order of $2\times10^{-3}$, but the scale is limited to less than $5\times10^{-4}$ to better show the evolution of $\Delta\varepsilon_{yy}$ in the intensity interval where most of the strain bursts occur for the two considered pixels P1 and P2.

\begin{figure}[t]
\begin{center}
\begin{center}
\includegraphics[width=8cm]{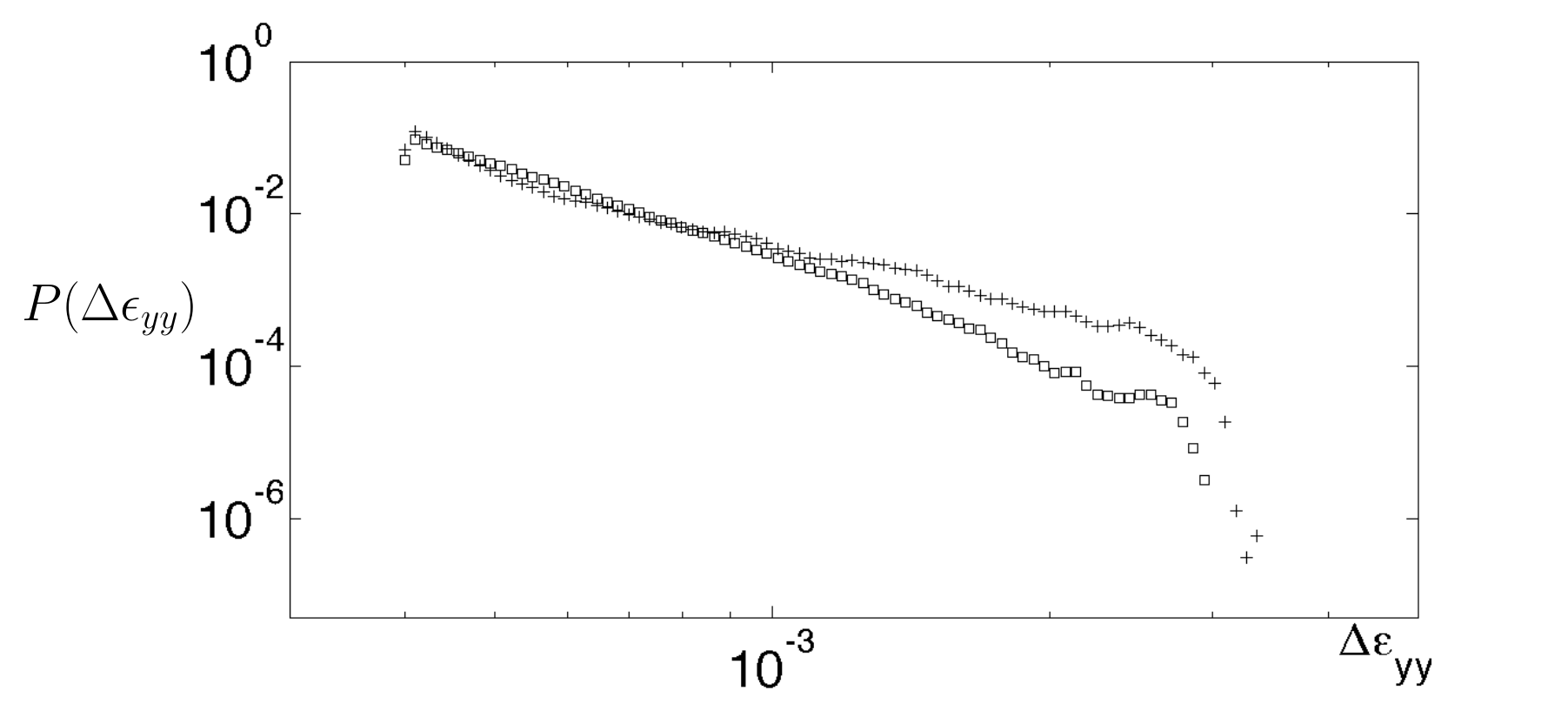}
\end{center}
\end{center}
\caption{Log-log plot of the probability distribution $P(\Delta\varepsilon_{yy})$ of the strain jumps $\Delta\varepsilon_{yy}$ (logarithmic binning) computed for all pixels in the sample during the forward (squares) and reverse (crosses) transformation.}
\label{single_pixel_statistics}
\end{figure}

To quantify statistically the degree of pixel-level burstiness in the strain-field evolution across the sample, we consider the amplitudes of $\Delta\varepsilon_{yy}$ recorded at all pixels along the entire transformation cycle. We threshold the values of $\Delta\varepsilon_{yy}$ at $4\times 10^{-4}$, to stay above the noise level of the local strain measurements (see the Appendix). It results that the pixel values of $\Delta\varepsilon_{yy}$ throughout the sample span about an order of magnitude, with heavy-tailed size distributions which are rather abruptly truncated, as shown in Fig.~\ref{single_pixel_statistics} separately for the forward and the reverse phase change. Each single pixel thus gives to the overall strain change a contribution that is bounded by the transformation strain between the austenitic and martensitic configurations. The large transformation events must thus necessarily exhibit some spatial structure, obtained as the sum of many smaller local events.

We now characterize the long-range aspects of the avalanche-mediated phase transformation. To take into account all the components of the evolving strain field, we consider the norm $|\Delta\varepsilon| =(\Delta\varepsilon_{yy}^2 +\Delta\varepsilon_{xx}^2 +2\Delta\varepsilon_{xy}^2)^{1/2}$ of the strain jumps rather than the sole $\Delta\varepsilon_{yy}$ component. We then check for any pixel P on which $|\Delta\varepsilon|$ is, at any time, higher than $4\times 10^{-4}$. Given any such P, we track a strain avalanche as the collection of pixels neighboring P whereon $|\Delta\varepsilon|$ is also above threshold. In this way, about 14,000 events are obtained from the strain evolution data collected along the hysteresis cycle. We recall that the time interval between images in the present tests is almost ten seconds (see the Appendix), which is many orders of magnitude larger than the temporal scales of avalanche durations as recorded for instance through acoustic emission \cite{PTZ_07}. Thus the avalanches evidenced here largely pertain only to a single image, and can give no indication on the actual time dynamics of the transformation, as each avalanche merges a large and variable number of bursty microscale events.

\begin{figure}[t]
\begin{center}
\begin{center}
\includegraphics[width=8cm]{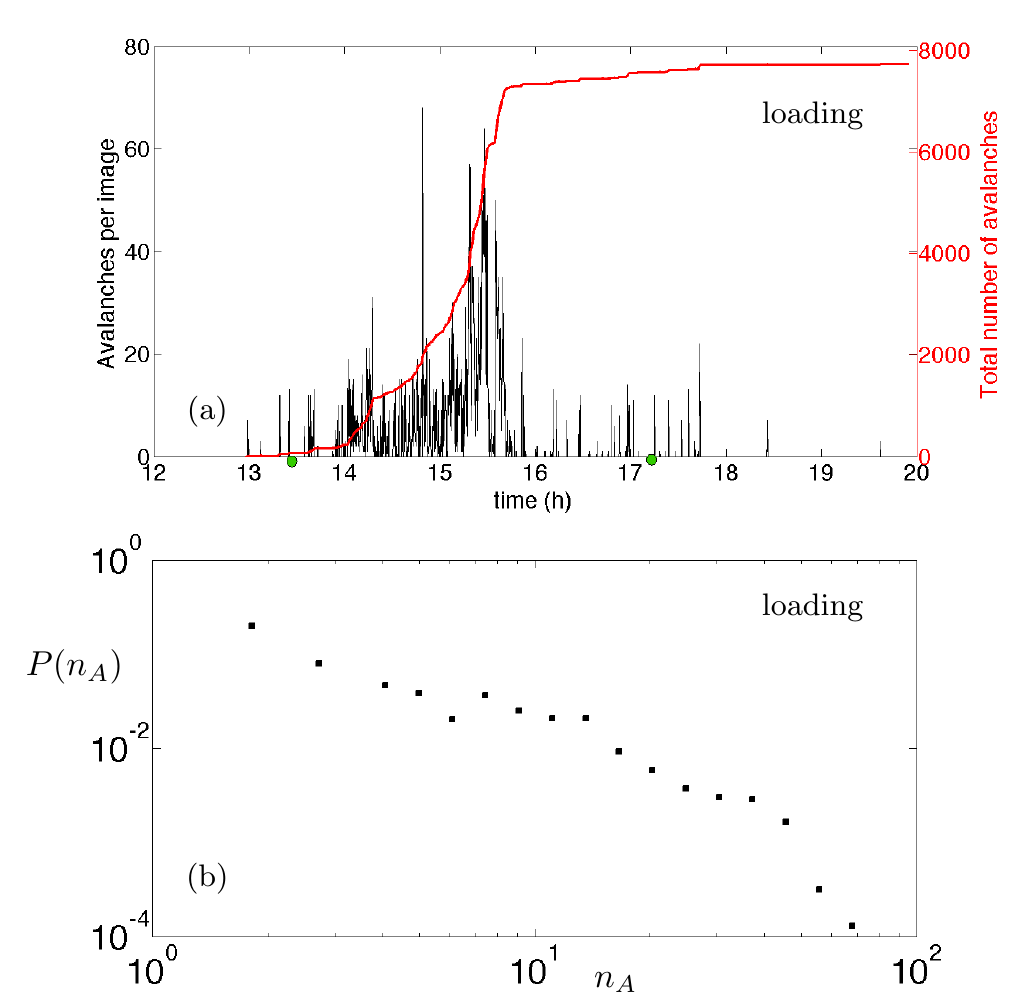}
\end{center}
\end{center}
\caption{Avalanching activity during the forward transformation. (a) Time evolution of the number  $n_A$ of avalanches detected per consecutive image (black) and its running sum (red). The green dots mark the reference points on the transformation strain plateaus. (b) Log-log plot of the corresponding probability distribution $P(n_A)$ (logarithmic binning).}
\label{avalanche_statistics}
\end{figure}

We first consider the number $n_A$ of avalanches detected per image in the sample. On loading, we record a total of almost 8,000 events, and about 6,000 events in the reverse transformation. In both case avalanching produces a non-stationary spiky signal, as shown for instance in Fig.~\ref{avalanche_statistics}(a) for the direct transformation. The corresponding heavy-tailed statistics in Fig.~\ref{avalanche_statistics}(b) evidence the very fluctuating character of these co-operative strain events, with $n_A$ spanning two orders of magnitude. The supplemental video~\#3~\cite{online_video} highlights the strain avalanching in the sample on loading, which helps visualize the bursty development of the martensite bands mentioned earlier. The supplemental video~\#4~\citep{online_video} shows the corresponding evolution of the strain avalanches, with their more complex microstructures, along the reverse transformation.

\begin{figure}[t]
\begin{center}
\includegraphics[width=9cm, height=13.5cm]{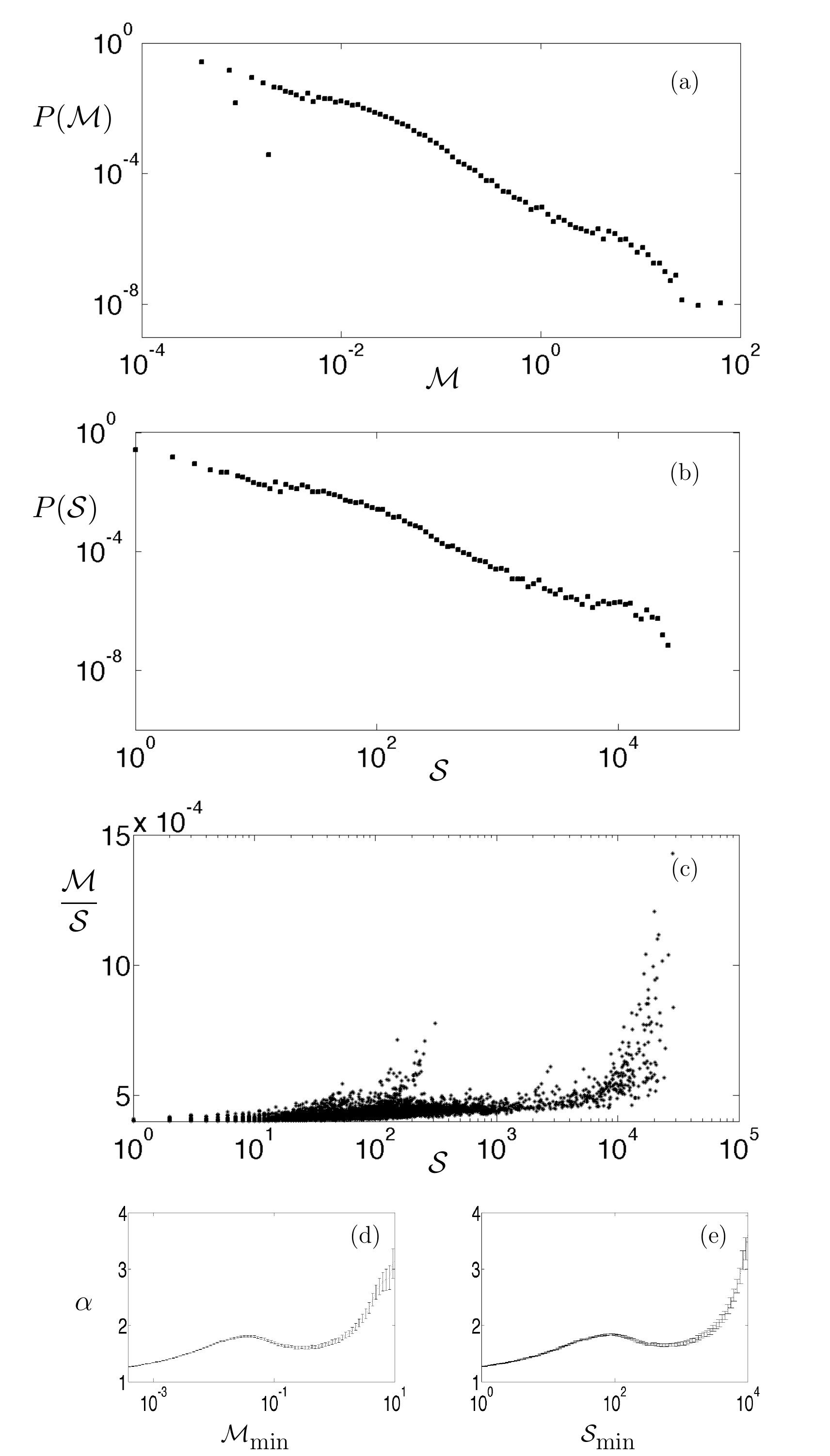}
\end{center}
\caption{Strain avalanching in the martensitic transformation. (a)-(b) Log-log plot of the probability distributions $P(\cal{M})$ and $P(\cal{S})$ of the strain avalanche magnitude $\cal{M}$ and the avalanche size $\cal{S}$ during the transformation cycle (logarithmic binning).
(c) Log-linear plot of the correlation between $\cal{M}$ and the ratio $\cal{M}/\cal{S}$ for the recorded strain avalanches. (d)-(e) Values of the exponents $\alpha$ respectively for $P(\cal{M})$ and $P(\cal{S})$, determined by the maximum likelihood method as a function of the lower cutoff imposed to the data.}
\label{pow_law}
\end{figure}

For each avalanche we then consider its \textit{size} $\cal{S}$ given by the total number of involved pixels, and its \textit{magnitude} $\cal{M}$ given by the integral of $|\Delta\varepsilon|$ over the avalanche. The quantity $\cal{M}$ measured for each strain avalanche gives the contribution of such event to the intermittent deformation detected in the sample at each consecutive image as the loading was applied, and, in particular, to the sample elongation detailed in Fig.~\ref{spikes_delta_epsilon_mean}. The probability distributions for both $\cal{S}$ and $\cal{M}$, computed over the full cycle are shown in Fig.~\ref{pow_law}. While none of the heavy-tailed statistics for strain intermittency considered so far (Figs.~\ref{spikes_delta_epsilon_mean}-\ref{avalanche_statistics}) seemingly exhibit power-law character, the log-log plots for $P(\cal{S})$ and $P(\cal{M})$ in Fig.~\ref{pow_law}(a)-(b) appear in remarkable agreement with the emergence of scale-free behavior during the transition process, which is well documented by the power-law size statistics for acoustic-emission events reported in many cases for the martensitic transformation of SMAs~\citep{Vives_94, vives95, Carrillo_Phys_Rev_B_97, Carrillo_Phys_Rev_Lett_98, perez_04a, Vives_Phys_Rev_B_09, niemann14, toth_14}.

We notice that the power-law character of the distributions of Figs.~\ref{pow_law}(b)-(c) is affected by a slightly larger frequency of events with sizes of the order of $S\sim 10^2$, which are mainly due to spurious avalanches due to grid defects. Also, Fig.~\ref{pow_law}(c) shows there is a positive correlation between $\cal{S}$ and $\cal{M}$, with a super-linear tendency for the avalanches with large $\cal{S}$ to have larger $\cal{M}$. This leads to an increase  (to about six) of the number of decades recorded for the avalanche magnitudes $\cal{M}$ in Fig.~\ref{pow_law}(a), as compared to those exhibited by the avalanche size $\cal{S}$ in Fig.~\ref{pow_law}(b). Furthermore, the defect-related events in the small plume of Fig.~\ref{pow_law}(c) for $S\sim 10^2$, widen the corresponding range of $\cal{M}$, and, at the corresponding scales in Figs.~\ref{pow_law}(a)-(b), locally slightly perturb both the distributions $P(\cal{S})$ and $P(\cal{M})$. This affects the stability of these distributions' exponents, which, as we see in Figs.~\ref{pow_law}(d)-(e), can both evaluated only as being in the range 1.5-1.8.  This is compatible with the exponent 1.6 recently proposed for the size statistics of the optically-observed microstructural transformation events recorded in~\citep{niemann14} at the surface of a Ni-Mn-Ga polycrystal undergoing a temperature-driven martensitic transformation.

\begin{figure}
\begin{center}
\includegraphics[width=\linewidth, height=2.8cm]{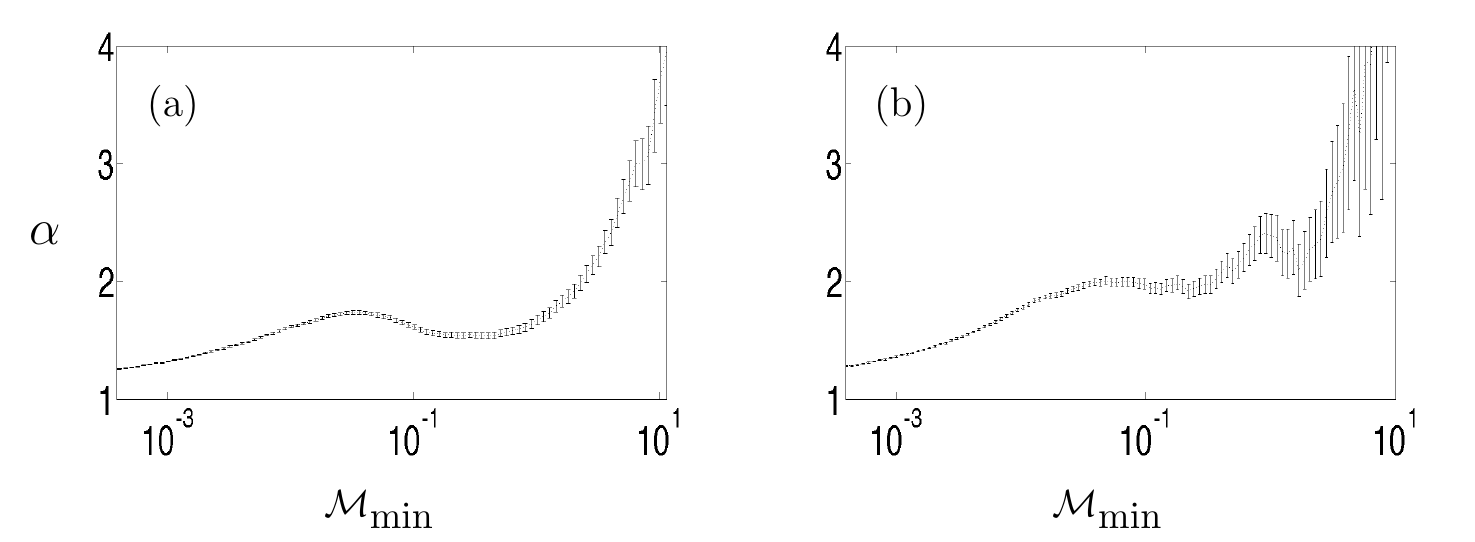}
\end{center}
\caption{Exponents $\alpha$ of the avalanches magnitude distribution $P(\cal{M})$ respectively for the forward (a), and for the reverse (b) transformation. The exponent values are determined by the maximum likelihood method as a function of the lower cutoff imposed to the data.}
\label{separate_exponents}
\end{figure}

\begin{figure}[htbp]
\begin{center}
\includegraphics[width=\linewidth]{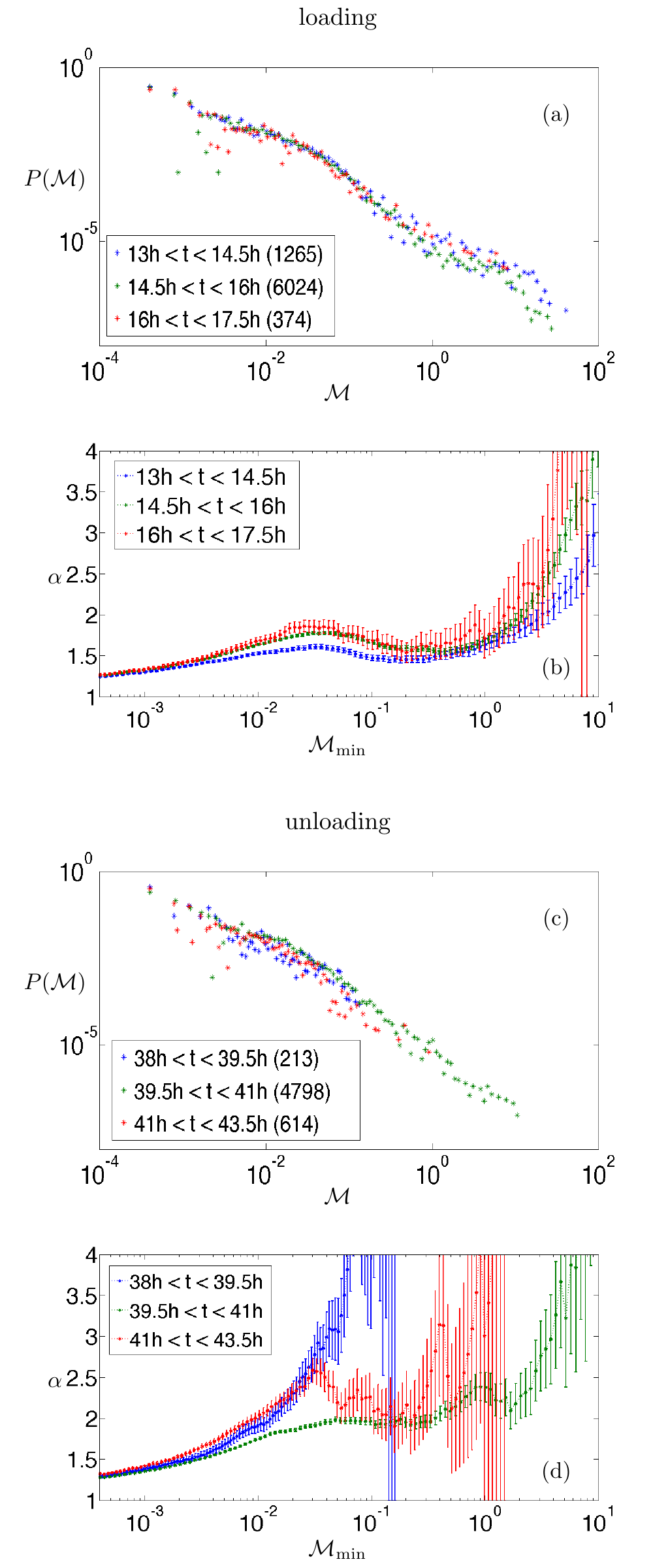}
\end{center}
\caption{(Colon online) Non-stationarity of the strain intermittency along the transformation cycle. (a) Probability distributions $P_i(\cal{M})$, $i = 1, 2, 3$ (logarithmic binning), of the avalanche magnitudes $\cal{M}$ during three consecutive equal-duration time intervals on the {\it forward}-transformation plateau. The inset indicates each interval, with, in parenthesis, the corresponding number of events:~1265, 6024, 374.
(b) Values of the exponent $\alpha$ for each distribution $P_i(\cal{M})$, determined by the maximum likelihood method as a function of the lower cutoff imposed to the data. (c-d) Data analogs of (a-b) for three consecutive equal-duration time intervals on the plateau for the {\it reverse} transformation, with corresponding number of events:~213, 4798, 614.
}
\label{non-stationarity}
\end{figure}

A further hysteretic aspect of the strain evolution in our material is manifested by the different exponents of the distributions $P(\cal{M})$ recorded for the strain avalanches on loading vs.~unloading, see Fig.~\ref{separate_exponents}. We notice that the exponent in the reverse transformation, near 2, is higher than the one on the forward transformation, near 1.5. An exponent asymmetry of this type was also recorded in \cite{toth_14} for the temperature-driven phase change in a NiMnGa SMA. Also acoustic emission data indicate a forward-reverse asymmetry in some SMA polycristals \cite{vives_rosinberg_12}.

We make a final brief analysis of the non-stationarity of the transformation process, already remarked from Figs.~\ref{spikes_delta_epsilon_mean} and \ref{avalanche_statistics}. We consider six consecutive time sub-intervals of equal duration over the transformation cycle (three for the forward and three for the reverse transformation). The number of events in each interval are indicated in the insets of Figs.~\label{non-stationarity}(a)-(c). We see that the non-stationarity of the process is marked by a much higher strain activity in the central portions of both plateaus, as opposed to their end intervals. This is seemingly analogous to the non-stationary signal recorded in other studies of SMA phase change \cite{niemann14}, or in the behavior of some hysteretic magnetic systems~\cite{durin06}. The present data also show there is a higher avalanching activity when the alloy is mostly austenitic than when martensite is predominant. The non-stationary character of the strain intermittency across the transformation is also evidenced through the probability distributions $P_i(\cal{M})$ of the avalanche magnitudes $\cal{M}$ pertaining to each one of the six time intervals, see Figs.~\ref{non-stationarity}(a) and (c). Figs.~\ref{non-stationarity}(b) and (d) show that the exponents of the $P_i(\cal{M})$ do not all have the same value, with the two distributions relative to the end intervals of the reverse transformation actually not possessing power-law character. As a further aspect of the asymmetry in the two phase change directions, we have from Figs.~\ref{spikes_delta_epsilon_mean} and~\ref{non-stationarity} that the recorded above-threshold avalanches not only are more numerous, but are also typically larger on loading vs.~unloading.

\section{Conclusions}
\label{section_5}

We have performed a full-field study, via the grid method, of the progress of the reversible cubic-monoclinic martensitic transformation in a Cu-Al-Be shape-memory single crystal. The phase change was driven by a stress-controlled loading device based on gravity capable of very small force increments, which applied to the sample a low-rate, monotonic load at largely isothermal conditions. We have thus recorded the stress-strain hysteresis in the forward and reverse transformation, and of the corresponding strain evolution in the microstructure, i.e.~in the spatial assemblies of the phases/variants in the sample. This was done by following the evolution in time of the surface strain maps in the sample obtained from the deformed grid data.

In general, due to the breaking of symmetry in the phase transition, the martensite exhibits multiple variants, with different orientations with respect to the parent lattice, and the detailed understanding of the microstructure is of wide interest in mechanics and materials science, as it is a key element influencing the macroscopic properties and behavior of a multiphase crystal \cite{nature_04, nature_14}. However, the design of the loading device sought to obtain for this study the simplest microstructures in a SMA single crystal, involving only the parent austenite and one variant of martensite. The very slow driving allowed us to highlight for the first time in a quantitative way the strain intermittency that marks the martensitic transformation in SMAs.

We have characterized the non-smooth progress of the transformation by first studying the intermittency for both the global (body-averaged) and local (pixel-level) strain measures. We have thereafter identified the organization of possibly long-range events, i.e.~strain avalanches, and studied their statistical properties (see also the supplementary videos \cite{online_video}). Many earlier experiments based on the acoustic emission accompanying the martensitic transformation in SMAs have shown some decades of avalanche-size scaling in the austenite-martensite dynamics \citep{Vives_94, vives95, Carrillo_Phys_Rev_B_97, Carrillo_Phys_Rev_Lett_98, perez_04a, Vives_Phys_Rev_B_09, niemann14, toth_14}. The statistics that we have now uncovered for our strain events indicate that power-law behavior pertains also to a broad distribution of strain-avalanche intensities, covering almost six decades well into the macroscopic sample scales. These results extend and make more quantitatively precise the results recently reported in \cite{niemann14} about the possible scale-free behavior of the surface optical activity in SMA martensitic transformation. We interpret this strain phenomenology as deriving from the evolution of the crystalline substance within a complex energy landscape whose local minimizers represent distinct  microstructural morphologies \cite{Ericksen75, Ball_James, book_PZ, Bhattacharya_03, CZ04, nature_04, PTZ_07}. The strain avalanches evidenced here mark the progress of the material through such minimizers under the loading \cite{PTZ_07}. The associated display of bursty scale-free behavior, as we presently observe, may confirm the indications about the critical nature of the reversible martensitic transformation in SMAs~\citep{Vives_94, vives95, Carrillo_Phys_Rev_B_97, Carrillo_Phys_Rev_Lett_98, Vives_Phys_Rev_B_09, niemann14, PTZ_07}, in analogy to what is observed also in the plastic flow of many crystals \cite{zapperi_carmen, zaiser, lev_umut}. We have also described a number of other aspects of the observed phase change, evidencing especially the non-stationarity of the material response along the transformation plateaus, and the asymmetric behavior characterized in the forward vs.~the reverse transformation process.

Further experimental work along the lines developed in the present study can be envisaged in various directions. This should advance our understanding, analysis, and possible control of phase coexistence and its evolution under changing external conditions, and may aid in the design of materials with targeted properties.

\medskip
{\bf Acknowledgements}. We thank E. Vives for several conversations. We acknowledge funding of the Italian PRIN Contract 200959L72B004. NB acknowledges support from the Italian Group of Mathematical Physics GNFM-INdAM.

\section*{Appendix}

The online Supplemental Material  \cite{online_video} accompanying this paper comprises four video files illustrating a number of different aspects of the phase transformation. Hereafter we give some details on the experimental methods.

Fig.~\ref{figure_1} shows schematically the Cu\,Al$_{11.4}$\,Be$_{0.5}$ (wt.~\%) single crystal sample and the testing apparatus. The sample dimensions are $17.78 \times 38 \times 0.94$ mm$^3$ respectively along the $x$-, $y$-, and $z$-directions in Fig.~\ref{figure_1}, the vertical distance between the clamped aluminum tabs being $38$ mm. The monitored gauge region has dimensions $17.78$ mm $\times$ 38 mm respectively along the $x$ and $y$ axes.

The rotation matrix giving the orientation of the austenitic cubic single crystal of Cu-Al-Be with respect to the reference axes was measured by X-ray diffraction, and is given by, in the same reference frame of Fig.~\ref{figure_1} (see also \cite{Delpueyo_Mech_Mater_12}):

\begin{equation}
R=\left[\begin{array}{lll}
0.7502 & 0.0570 & 0.6588\\
-0.6612 & 0.0721 & 0.7467\\
-0.0049 & -0.9958 & 0.0918
\end{array}\right].
\end{equation}

We measured the in-plane displacement, the linear strain components and the local rotation on the gauge region of the sample through the grid method. This technique derives these quantities from the images of a grid captured by a camera as the sample deforms under the loading. Grids are transferred onto the sample surface using the procedure described in~\citep{Piro_Exp_Tech_04}. The grid pattern (square, with pitch $p = 0.2$~mm) was printed on a thin polymeric substrate using a 12,000-dpi high-definition printer and then laid on a thin adhesive layer (E$504$ glue from Epotecny, whose white color optimizes the visual contrast with the black grid lines. After curing for about $40$\,h at $37\,^\circ$C, the polymeric substrate is carefully removed, but the black ink of the grid lines remain bonded on the thin adhesive layer and the sample becomes ready for testing. The deformation of the grid can reasonably be assumed to be the same as that of the surface of the sample as the glue layer is very thin (some tenths of mm), and has negligible stiffness compared to that of the sample, making this a non-intrusive measurement technique. The maximum strain that can be sustained by the grid without debonding or cracking is about $18\%$, which can well accommodate any strain reached during the martensitic transformation in our alloy.

In our experiment, the images of the deformed grid during the tests were captured by a Sensicam QE camera featuring a $12$-bit/$1040\times1376$ pixel sensor and a $105$-mm Sigma lens. The camera was secured on a mounting plate whose position was adjusted so that the lines of the grid were parallel to the pixels of the sensor along both the horizontal and vertical axes. An appropriate magnification allowed to clearly distinguish the black lines of the grid. This magnification was adjusted so that one grid pitch was encoded with 5~pixels. Three movable light guides fed by a KL $2500$ LCD cold light source were used during image acquisition to obtain a nearly-uniform lighting of the grid. About 20,000 images of the deforming sample were obtained over the whole duration of the test, one about every 8 sec, with 15-min pauses about every 100\,min for data recording and refilling the water reservoirs which feed the pumps.

As the grids are the superimposition of two perpendicular straight line patterns, each of them can be considered as a periodic marking, whose change due to the deformation during the test is proportional to the sought in-plane displacement~\citep{Surrel_00}:
\begin{equation}
\label{eq:delta}
   u_x=-\dfrac{p}{2\pi}\;\Delta\phi_x \vspace{+3mm}, \quad    u_y=-\dfrac{p}{2\pi}\;\Delta\phi_y,
\end{equation}
where $\phi_x$ and $\phi_y$ are the local phases along $x$ and $y$, respectively. In practice, $\phi_x$ and $\phi_y$ are deduced from the grid images by calculating at any pixel the windowed Fourier transform (WFT), see~\citep{Surrel_00}, for which a Gaussian window \citep{Badulescu_Meas_Sci_Technol_09} was employed in the present study. The possible appearance of local fictitious strains induced by slight grid printing defects was strongly reduced by using a motion compensation technique described in \cite{Badulescu_Meas_Sci_Technol_09}.

The above measurement and data-treatment methods provide strain maps with a good compromise between strain resolution and spatial resolution, a key factor for distinguishing details in the strain maps. The spatial resolution of the technique, i.e.~the shortest distance between two independent measurements, is considered here to the equal to the width of the Gaussian envelope used in the WFT, which we consider equal to $6\,\sigma$ according to the classical 3-$\sigma$ rule. As we set $\sigma = p$, which is the smallest value that can be chosen for $\sigma$~\cite{Sur_Inverse_Problems_Imaging_14}, and as 5~pixels encode one grid pitch~$p$, we obtain a spatial resolution equal to $30$ pixels in this study, that is, 1.2~mm on the sample. Random noise propagating from the camera sensor is one of the main sources of disturbance in the strain maps. Considering only the latter, and taking into account that the camera was set to average 128 frames to provide one grid image, a threshold equal to $4\times10^{-4}$ for  $|\Delta\varepsilon|$ has been estimated \cite{Grediac_Strain_14, ?15SurGr}, and has been used to separate the avalanches within the noise floor in Sect.~\ref{section_4}. �
The same value has also been considered for the strain components $\varepsilon_{xx}$, $\varepsilon_{yy}$ and $\varepsilon_{xy}$. Correspondingly, for the average strain components and their variations, such as $\overline{\varepsilon}_{yy}$, $\Delta\varepsilon_{yy}$, etc., computed over a whole strain map, we have a lower threshold, as we rely on ~$4\times10^{5}$ measurements to calculate a single global quantity for each field. The obtained value was then rounded up to be equal $1\times10^{-6}$, which was used as threshold for the analysis in Sect.~\ref{section_2}. The strain-increment maps show in some cases the presence of parasitic fringes, probably originating from a slight moir\'e effect between grid and camera sensor. Such quasi-periodic noise, for which there is at present no automatic elimination procedure available in the case of strain distributions \citep{Grediac_15a, Sur_15c}, may shade some actual strain-map features, thus spuriously increasing the number of apparently independent strain avalanches. This undesirable effect constitutes one of the limits of the present analysis.
We have however heuristically performed avalanche identification also by using thresholds for $|\Delta\varepsilon|$ lower than the value $4\times10^{-4}$ pertaining to the sole sensor noise. We checked directly that this only minimally affects the statistical results reported in Figs.~10-13, confirming their robustness.

\bibliographystyle{elsarticle-harv}

\end{document}